\documentclass{raa}            

\usepackage{graphicx,times}             

\begin{document}

  \title{The timing behavior of magnetar Swift J1822.3$-$1606: timing noise or
   a decreasing period derivative?
}

   \volnopage{Vol.0 (200x) No.0, 000--000}      
   \setcounter{page}{1}          

     \author{H. Tong
      \inst{1}
   \and R. X. Xu
      \inst{2}
   }

   \institute{Xinjiang Astronomical Observatory, Chinese Academy of Sciences, Urumqi, Xinjiang 830011,
    China; {\it tonghao@xao.ac.cn}\\
        \and
             KIAA and School of Physics, Peking University, Beijing 100871, China\\
           }

   \date{Received~~2012 month day; accepted~~2012~~month day}

\abstract{ The different timing results of the magnetar Swift
J1822.3$-$1606 is analyzed and understood theoretically. It is
pointed that different timing solutions are caused not only by
timing noise, but also that the period derivative is decreasing
after outburst. Both the decreasing period derivative and the large
timing noise may be originated from wind braking of the magnetar.
Future timing of Swift J1822.3$-$1606 will help us make clear
whether its period derivative is decreasing with time or not.
 }

   \authorrunning{Tong \& Xu}            
   \titlerunning{Comments on the timing of magnetar Swift J1822.3$-$1606}  

   \maketitle

%
%
\section{Introduction}           

Magnetars are peculiar pulsar-like objects. They are assumed to be
neutron stars powered by strong magnetic field decay (Duncan \&
Thompson 1992). A neutron star is often confirmed as a magnetar if
its surface dipole magnetic field is higher than the quantum
critical field ($B_{\rm QED}=4.4\times 10^{13} \,\rm G$). The
surface dipole magnetic field is calculated from the period and
period derivative (assuming magnetic dipole braking, Kouveliotou et
al. 1998). However, the magnetic dipole braking assumption will also
provide challenges to the magnetar model. One example is the
existence of the low magnetic field magnetar (Rea et al. 2010; Tong
\& Xu 2012). Alternatively, it is possible that magnetars are wind
braking (Tong et al. 2013 and references therein). Wind braking
would help us to explain the controversial timing results
of the magnetar Swift J1822.3$-$1606.

Swift J1822.3$-$1606 is a magnetar candidate, discovered by {\it Swift}/BAT on 2011 July 14
(Cummings et al. 2011). Up to now, different timing results are obtained for this source
(Livingstone et al. 2011; Rea et al. 2012; Scholz et al. 2012). The reported period derivative
differs by a factor about three. The corresponding characteristic magnetic field can be larger or
smaller than the quantum critical field.
This is directly related to whether this source is another low magnetic field magnetar or not.

In Rea et al. (2012) and Scholz et al. (2012) observational papers,
they mainly discussed the timing noise effect. In their opinion, it
is the the large timing noise that results in different period
derivative measurements of Swift J1822.3$-$1606. In this paper, we
explore another effect. The period derivative of Swift
J1822.3$-$1606 may be decreasing with time. Therefore, it is natural
that different period derivatives are obtained using different data
sets. The physical reason may be that magnetars are wind braking
(Tong et al. 2013). A decaying particle wind after outburst will
result in a decreasing period derivative.

Model description and quantitative calculations are presented in Section 2.
Discussions and conclusions are presented in Section 3.

\section{Modeling the spin down rate of Swift J1822.3$-$1606}

\subsection{Description of observations and theory}

In Rea et al. (2012), they reported two period derivatives of Swift J1822.3$-$1606 (Section 3.2 there).
Using the first 90 days observations,
a period derivative $\dot{P}=1.6(4)\times 10^{-13}$ is obtained (the last digit uncertainties are at 1 $\sigma$
confidence level). Considering the whole 275 days data,
the corresponding period derivative is $\dot{P}=0.83(2)\times 10^{-13}$.
These two values provide some hints that the period derivative is decreasing with time.
The large uncertainty in short time data set may be caused by timing noise.
Similar behavior can also be seen in Livingstone et al. (2011) and Scholz et al. (2012).
In Livingstone et al. (2011), using 84 days observations, a period derivative
$\dot{P}=2.55(22)\times 10^{-13}$ is reported. Using 402 days observations, Scholz et al. (2012)
reported three solutions of period derivatives:
$\dot{P}=0.683(21)\times 10^{-13}$ (fitting with period and period derivative),
$\dot{P}=1.71(7)\times 10^{-13}$ (fitting with period and two period derivatives),
$\dot{P}=3.06(21)\times 10^{-13}$ (fitting with period and three period derivatives).

Similar behaviors are also seen in other magnetars. Since the early stage of magnetar timing
studies, it is found that magnetars have a higher level of timing noise than normal pulsars
(Gavriil \& Kaspi 2002; Woods et al. 2002). Large period derivative variations are seen
in AXP 1E 2259+586 (Kaspi et al. 2003), AXP 1E 1048.1$-$5937 (Gavriil \& Kaspi 2004),
SGR 1806$-$20 (Woods et al. 2007), and AXP 1E 1547.0$-$5408 (Camilo et al. 2008).
Two neat examples are AXP XTE J1810$-$197 (Camilo et al. 2007) and the
radio loud magnetar PSR J1622$-$4950 (Levin et al. 2012). In these two sources,
a decreasing period derivative is observed while the star's X-ray luminosity is decreasing
after outburst. Therfore, from previous observations, there may also be
large timing noise in Swift J1822.3$-$1606. At same time, its period derivative may also
decrease with time (a decreasing X-ray luminosity is also observed).
This may explain why a lower period derivative is obtained when using longer
time span of observations.

The physics for a varying period derivative may be that magnetars are wind braking
(Tong et al. 2013). The decay of strong magnetic field will power the star's X-ray luminosity.
At the same time, a (magnetism-powered) particle wind is also generated. The rotational energy
of magnetars is mainly carried away by this particle wind. A varying particle wind
naturally results in a varying period derivative. The fluctuations of this particle wind may
account for the large timing noise in magnetars. Since both the X-ray luminosity and
the particle wind luminosity are from magnetic field decay, a model independent estimate of the
particle wind luminosity is $L_{\rm p} \sim L_{\rm x}$, where $L_{\rm p}$ and $L_{\rm x}$
are the particle wind luminosity and the X-ray luminosity, respectively. The origin for this
particle wind may be either internal (e.g., low amplitude seismic activities, Thompson \& Duancan 1996),
or magnetospheric (e.g., coronal particles, Beloborodov \& Thompson 2007). For details of
wind braking of magnetars and discussion of other models, see Tong et al. (2012) and references
therein.

\subsection{Calculations for Swift J1822.3$-$1606}

X-ray observations of Swift J1822.3$-$1606 have given its flux evolution with time.
Using the flux evolution function and its extrapolations, we can calculate the theoretical
period derivative as a function of time. The longest time span of X-ray observations of Swift J1822.3$-$1606
is done by Scholz et al. (2012, 400 days observations). According to Scholz et al.
(2012), a double exponential flux decay model is prefered.
\begin{equation}\label{F(t)}
F(t) = F_1 \exp[ -t/\tau_1] +F_2 \exp[ -t/\tau_2] + F_{\rm q},
\end{equation}
where $F(t)$ is the 1$-$10 $\rm keV$ source flux as a function of time, $t$ is in units of days
since BAT trigger time (MJD 55756.5), $\tau_1=15.5 \,\rm days$ and $\tau_2 =177 \,\rm days$
are the two decay time scales, $F_1=20.9 \times 10^{-11} \,\rm erg \, cm^{-2} \,s^{-1}$
and $F_2=1.74 \times 10^{-11} \,\rm erg \, cm^{-2} \,s^{-1}$ are the two flux normalizations,
$F_{\rm q} =3\times 10^{-3} \times 10^{-11} \,\rm erg \, cm^{-2} \,s^{-1}$ is the fixed quiescent
flux (constrained by {\it ROSAT}). See Scholz et al. (2012, Section 3.3 there) for details.

The rotational energy loss rate due to an isotropic particle wind is
proportional to $L_{\rm p}^{1/2}$ (Section 3 in Tong et al. 2013).
Therefore, the period derivative will evolve with time as (short
term evolution, e.g. several years) $\dot{P}(t) \propto L_{\rm
p}^{1/2} \propto L_{\rm x}^{1/2} \propto F(t)^{1/2}$. Including a
constant factor
\begin{equation}
\dot{P}(t) = N_0 \, F(t)^{1/2},
\end{equation}
where $N_0$ is the normalization constant. The observational period derivative
is the average value over a certain time span. Expanding the period at epoch $t_1$,
\begin{equation}\label{P expansion}
P(t) = P(t_1) + \dot{P}(t_1)(t-t_1),
\end{equation}
where $P(t)$ and $P(t_1)$ are the rotation period at time $t$ and $t_1$, respectively,
$\dot{P}(t_1)$ is the period derivative at $t_1$. Therefore, the observational 
period derivative for time span $t-t_1$ is ($t$ is the end time, $t_1$ is the starting time)
\begin{equation}
\dot{P}_{\rm obs} (t-t_1) =\dot{P}(t_1) = \frac{1}{t-t_1} (P(t)-P(t_1)).
\end{equation}
Rewriting the above equation,
\begin{eqnarray}
\dot{P}_{\rm obs} (t-t_1) &=& \frac{1}{t-t_1} \int_{t_1}^{t} \dot{P}(t^{\prime}) {\rm d} t^{\prime}\\
&=& N_0 \frac{1}{t-t_1} \int_{t_1}^{t} F(t^{\prime})^{1/2} {\rm d} t^{\prime}\\
&=& N_0 g(t,t_1)\label{Pdotobs},
\end{eqnarray}
where $g(t,t_1)=\int_{t_1}^{t} F(t^{\prime})^{1/2} {\rm d} t^{\prime}/(t-t_1)$, $F(t)$ is obtained
by fitting the observational flux decay (equation (\ref{F(t)})).

The timing of Livingstone et al. (2011) is done for time span $85.5-1.5$ days since BAT trigger
time. While the timing of Scholz et al. (2012) is for time span $404.5-2.5$ days since BAT trigger time.
According to equation (\ref{Pdotobs}), the ratio of period derivative between Scholz et al. (2012)
and Livingstone et al. (2011) should be $g(404.5,2.5)/g(85.5,1.5)=0.48$. The observational value
is $0.683(21)/2.55(22)$, for solution 1 (The case of solution 2 and solution 3 will be discussed 
in the discussion section). 
The observation and theory are consistent within uncertainties. The same can also be done
for the timing of Rea et al. (2012). Using the observational flux decay there, the theoretical period
derivative ratio between 275 days and 90 days observation is $0.60$. While the observational value
is $0.83(2)/1.6(4)$. The observation and theory are consistent with each other.

\begin{figure}
\centering
  \includegraphics[width=0.7\textwidth]{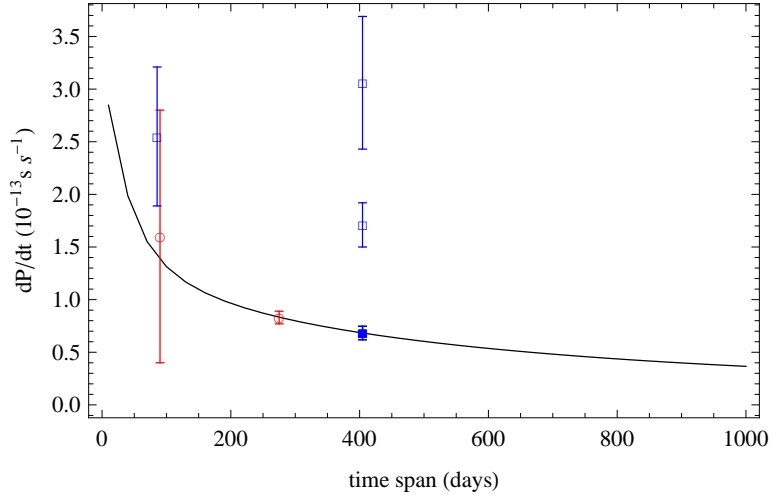}\\
  \caption{Theoretical period derivative as a function of observational time span.
  The continuous line is the theoretical period derivative. Red circles are timing data from
  Rea et al. (2012). Blue squares are timing data from Livingstone et al. (2011)
  and Scholz et al. (2012). The filled blue square is taken as normalization
  of the theoretical curve. The error bars are 3 $\sigma$. }\label{gpdot}
\end{figure}

We can also plot the theoretical period derivative as a function of time span. Employing the period
derivative of solution 1 in Scholz et al. (2012) as the normalization,
the predicted period derivative as a function of time is
\begin{equation}
\dot{P}_{\rm obs}(t-t_1) = \dot{P}_{\rm obs}(404.5-2.5) \frac{g(t,t_1)}{g(404.5,2.5)}.
\end{equation}
The timing solutions in Livingstone et al. (2011), Rea et al. (2012), Scholz et al. (2012)
are taken at different epoch (i.e., different $t_1$). However, the differences are only one
or two days. Therefore, this difference is negligible. $t_1=2.5$ is assumed in the following
calculations (the $t_1$ value in Scholz et al. (2012)). Figure \ref{gpdot} shows the theoretical
period derivative and the current observational data. The theoretical curve (using solution 1 in Scholz et al.
(2012) as normalization) is consistent with the timing of Rea et al. (2012). The large uncertainties
in the timing of Livingstone et al. (2011) and 90 days timing result of Rea et al. (2012)
may be due to timing noise.

In the future, when longer time span observations are available, a smaller period derivative is expected.
For example, 800 days of timing observations will result in a period derivative
$\dot{P}=0.44 \times 10^{-13}$. This is the theoretical period derivative averaged over 800 days.
If separate timing can be done for the early 400 days and the late 400 days, a smaller period
derivative is expected. Current 400 days timing gives a period derivative $\dot{P} =0.683 \times 10^{-13}$.
A period derivative $\dot{P} = 0.19 \times 10^{-13}$ is expected for the late 400 days timing only.
It is about three times smaller. Future timing observations of Swift J1822.3$-$1606 will help
us make clear whether its period derivative is decreasing with time or not.

\section{Discussions and conclusions}

The above calculations are mainly base on equation (\ref{P
expansion}). In equation (\ref{P expansion}), only the first period
derivative is included in the expansion. The observed $\dot{P}$ is
the average value of period derivative over the observational time
span. During timing studies higher order period derivatives may also
be included (e.g., solution 2 and solution 3 in Scholz et al. 2012).
When higher order period derivatives are considered, the
corresponding $\dot{P}$ will approach its instantaneous value at the
expansion epoch. Therefore, the reported $\dot{P}$ represents
earlier value when higher order period derivatives are included. If
the physical spin down rate is decreasing with time, we should see a
larger $\dot{P}$ when higher order period derivatives are included.
This is just the three timing solutions in Scholz et al. (2012).
Therefore, the three timing solutions of Scholz et al. (2012)
provide us another evidence that the period derivative of Swift
J1822.3$-$1606 is decreasing with time.

When calculating the theoretical spin down rate, the particle wind luminosity is assumed to equal to
the soft X-ray luminosity. The actual wind luminosity may have a more or less different value. After the ourburst,
the star's X-ray luminosity decreases with time. Since the particle wind is also from magnetic field decay,
then it is natural that the wind luminosity also decreases with time. Therefore, a decreasing
period derivative is always expected irrespective of the details of particle wind luminosity.
In the long term run,
the X-ray luminosity will return to its quiescent value. The particle wind will also relax to its
quiescent state. The long term predicted period derivative is very sensitive to the condition of the
quiescent state. When assuming $L_{\rm p}=L_{\rm x}$, the period derivative at late time will be
$\dot{P} \propto F_{\rm q}^{1/2}$, where $F_{\rm q}$ is the quiescent flux. A quiescent flux ten
times higher, the late time period derivative will be be three times larger.

The surface dipole field obtained by assuming magnetic dipole braking is only the effective field strength.
In the presence of strong particle wind, the rotational energy loss rate is amplified. For a given
period derivative, the resulting dipole field will be much lower (Tong et al. 2013). In the actual case,
the geometry (e.g., the magnetic inclination angle) will also affect the spin down history of the neutron star
(Tong \& Xu 2012). In the case of normal pulsars, the magnetic dipole braking assumption is a not too
bad lowest order approximation (Xu \& Qiao 2001). However, in the case of magnetars, the magnetic dipole braking
assumption will be too simple even to the lowest order approximation. An alternative is that magnetars are
wind braking (Tong et al. 2013; Tong, Yuan \& Liu 2013). A decaying particle wind can result in a decreasing
period derivative of Swift J1822.3$-$1606.

Another explanation for the decreasing period derivative is the twisted magnetosphere model
(Thompson et al. 2002; Beloborodov 2009). After the outburst, the magnetar magnetosphere gradually untwists.
Therefore, the effect dipole magnetic field will decrease. This will cause a decreasing period derivative.
However, the twisted magnetosphere model may have difficulties in explaining the short time scale
period derivative variations (Camilo et al. 2007; Levin et a. 2012). In the above wind braking of magnetars,
the wind luminosity can vary dramatically on short time scales.
Such difficulties no longer exist in the wind braking model.

In conclusion, the different timing results of Swift J1822.3$-$1606
are caused not only by its timing noise, but also by its decreasing
period derivative. The decreasing period derivative and large time
noise may be both originated from wind braking. Future timing
observations of Swift J1822.3$-$1606 will help to make clear whether
its period derivative is decreasing with time or not. This would
also help us to answer whether wind braking is important in this
source or not.

\section*{Acknowledgments}

The authors would like to thank P. Scholz for providing the
observational value of $F_1$ and $F_2$ and for explanations; and S.
Dai, Y. Xie for discussions. H. Tong would like to thank KIAA at PKU
for visiting support. This work is supported by National
Basic Research Program of China (2012CB821800, 2009CB824800),
National Natural Science Foundation of China (11103021, 11225314,
10935001), West Light Foundation of CAS (LHXZ201201), Xinjiang Bairen project, 
and Qing Cu Hui of CAS.

\label{lastpage}

\end{document}